\documentclass[sigconf,authorversion,nonacm]{acmart}

\usepackage{makecell}
\usepackage{subcaption}
\usepackage{algorithm}
\usepackage{algorithmic}
\usepackage{threeparttable}
\usepackage{stfloats}
\newcommand*\rfrac[2]{{}^{#1}\!/_{#2}}

\newcounter{noteXPctr} 

\AtBeginDocument{%
  \providecommand\BibTeX{{%
    \normalfont B\kern-0.5em{\scshape i\kern-0.25em b}\kern-0.8em\TeX}}}

\acmSubmissionID{123-A56-BU3}

\begin{document}

\title{Learning to Learn Financial Networks for Optimising Momentum Strategies}


\author{Xingyue (Stacy) Pu}
\email{xingyue.pu@eng.ox.ac.uk}
\affiliation{
  \institution{Oxford-Man Institute of Quantitative Finance}
  \institution{University of Oxford}
  \city{Oxford}
  \country{UK}
}

\author{Stefan Zohren}
\email{stefan.zohren@eng.ox.ac.uk}
\affiliation{
  \institution{Oxford-Man Institute of Quantitative Finance}
  \institution{University of Oxford}
  \city{Oxford}
  \country{UK}
}

\author{Stephen Roberts}
\email{sjrob@robots.ox.ac.uk}
\affiliation{
  \institution{Oxford-Man Institute of Quantitative Finance}
  \institution{University of Oxford}
  \city{Oxford}
  \country{UK}
}

\author{Xiaowen Dong}
\email{xdong@robots.ox.ac.uk}
\affiliation{
  \institution{Oxford-Man Institute of Quantitative Finance}
  \institution{University of Oxford}
  \city{Oxford}
  \country{UK}
}


\begin{abstract}

Network momentum provides a novel type of risk premium, which exploits the interconnections among assets in a financial network to predict future returns. However, the current process of constructing financial networks relies heavily on expensive databases and financial expertise, limiting accessibility for small-sized and academic institutions. Furthermore, the traditional approach treats network construction and portfolio optimisation as separate tasks, potentially hindering optimal portfolio performance. To address these challenges, we propose L2GMOM, an end-to-end machine learning framework that simultaneously learns financial networks and optimises trading signals for network momentum strategies. The model of L2GMOM is a neural network with a highly interpretable forward propagation architecture, which is derived from algorithm unrolling. The L2GMOM is flexible and can be trained with diverse loss functions for portfolio performance, e.g. the negative Sharpe ratio. Backtesting on 64 continuous future contracts demonstrates a significant improvement in portfolio profitability and risk control, with a Sharpe ratio of 1.74 across a 20-year period. 


\end{abstract}



\maketitle

\section{Introduction}

Financial networks represent interconnections among financial entities such as institutions, markets and assets, and are constructed to facilitate the analysis of various downstream tasks. These tasks primarily focus on systemic risk assessment and market structure analysis, and more recently on alpha research and portfolio optimisation \cite{aliSharedAnalystCoverage2020,leeTechnologicalLinksPredictable2019}. Network momentum is a type of network alpha signals, which is based on the idea that the future return of one company in the financial network can be predicted from the past returns of companies to which it is connected, a phenomenon known as momentum spillover \cite{aliSharedAnalystCoverage2020}. \par

The construction of financial networks is crucial to the success of network momentum signals. Nodes often represent individual stocks, while edges capture the relations between them and are constructed based on various financial or economic factors. These factors include, but are not limited to, industry or business similarity \cite{moskowitzIndustriesExplainMomentum1999, boniAnalystsIndustriesPrice2006}, geographical proximity \cite{parsonsGeographicLeadLagEffects2020, korniotisStateLevelBusinessCycles2013a} news co-mentions \cite{wanSentimentCorrelationFinancial2021} and supply-demand links \cite{cohenEconomicLinksPredictable2008b, menzlyMarketSegmentationCrosspredictability2010}. However, the construction of these networks often relies on acquiring expensive databases from domain experts with deep understanding of finance and economics, making it costly for small-sized, non-profit or academic institutions.
Alternatively, we may adopt the reasonable assumption that momentum spillover is subtly echoed in the historical pricing data, which are more readily accessible. Therefore, if an effective model could be developed to learn the relationship between companies from readily available data, it could significantly lessen the dependence on costly data acquisition and specialised financial knowledge. Nevertheless, creating such a model remains a significant open challenge in the field of machine learning and quantitative finance. 

Another limitation of the literature is that the construction of financial networks or graphs\footnote{In this paper, we use \textit{graph} and \textit{network} interchangeably, and \textit{neighbours} and \textit{peers} interchangeably to represent the assets one is connected to. In the literature, the connected assets are also sometimes referred to as \textit{linked} assets. 
} is often detached from the subsequent downstream tasks. For example, analysts tend to focus more on economic justification when constructing these graphs, giving less consideration to whether they would lead to optimal portfolio performance. 
However, it is possible that the method of graph construction, such as deducing a custom sales ratio formula to determine the connection strength between two companies \cite{cohenEconomicLinksPredictable2008b}, may not help with achieving optimal portfolio performance. In other words, the graph may not be effectively optimised with respect to portfolio performance. This underscores the need for a task-specific graph construction framework. \par

In response to these considerations, we propose an end-to-end machine learning framework capable of simultaneously learning financial graphs and optimising the trading signals from momentum features built from these networks.  Our framework leverages the concept of \textit{learning to learn graph topologies} (L2G) \cite{puLearningLearnGraph2021}. Traditional graph learning methods typically employ convex optimisation to solve for an optimal graph given observed data on nodes \cite{dongLearningGraphsData2019, dongLearningLaplacianMatrix2016}. L2G reformulated this process into learning a parametric mapping, where node features are input and the graph serves as the output, by unrolling the optimisation algorithm into a forward propagation of a neural network. 
Taking advantage of the flexibility in neural architecture design and training, 
we propose to incorporate into L2G an additional layer for constructing network momentum features, and train the model with loss being portfolio performance metrics in an end-to-end fashion. \par

We test the proposed strategy, \textit{Learning to Learn Financial Networks for Optimising Momentum Strategies} (L2GMOM), on 64 continuous future contracts, spanning commodities, equities, fixed income funds, and foreign currencies. We obtain a Sharpe ratio of 1.74 across 20-year backtest period from 2000 to 2020
\cite{bazDissectingInvestmentStrategies2015, limEnhancingTimeSeries2020}. In addition, the learned financial graphs 
effectively elucidate the momentum spillover for portfolio construction, which enhances the interpretability of the proposed methods. Our model also excels in terms of computational efficiency, offering rapid inference in milliseconds and bypassing the need for daily graph optimisation. 


The major contributions of our work are threefold: Firstly, we present a novel methodology that uses easily accessible pricing data to infer financial networks between assets. This approach eliminates the costly requirement for alternative datasets and reduces human bias. Secondly, we propose an end-to-end learning framework that is capable of both inferring financial networks and simultaneously optimising the downstream task of portfolio construction. This ensures the financial networks are directly optimised in relation to portfolio performance. Finally, our proposed methods demonstrate strong empirical portfolio performance in both profitability and risk control.
 
\section{Preliminary}

\subsection{Individual Momentum Strategy}


The concept of individual asset momentum, proposed by \cite{moskowitzIndustriesExplainMomentum1999}, explores the profitability by longing past winner assets and shorting past underperformers. A time-series (TSMOM) momentum can be therefore constructed, with the daily return defined as
\begin{equation}
\label{eq: TSMOM}
r_{t:t+1}^{\text{portfolio}} := \frac{1}{N_t}\sum_{i=1}^{N_t} x_{i,t} \frac{\sigma_{\text{tgt}}}{\sigma_{i,t}} r_{i, t:t+1}
\end{equation}%
where $N_t$ is the number of assets at day $t$, $r_{i, t:t+1}$ is the daily return of asset $i$. With the target annualised volatility $\sigma_{\text{tgt}}$, the asset return is scaled by its annualised realised volatility $\sigma_{i,t}$. Here, $\sigma_{\text{tgt}} = 0.15$ following literature standards \cite{limEnhancingTimeSeries2020}, and $\sigma_{i,t}$ is estimated from an exponential weighted moving standard deviation with a 60-day span on daily returns. The position, or trading signal $x_{i,t} \in [-1,1]$ at day $t$ for asset $i$, specifies the strategy. The core of momentum strategies is the construction of the trading signal $x_{i,t}$. It often consists of two steps - trend estimation and position sizing. We illustrate this with the following examples. \par


\paragraph{Annual Return} \citet {moskowitzTimeSeriesMomentum2012} proposed to use past 1-year asset return $r_{i, t-252:t}$ as a trend estimation, and the direction as the position $x_{i,t} = \text{sgn}(r_{i, t-252:t})$.

\paragraph{MACD} \citet{bazDissectingInvestmentStrategies2015} constructed trend estimation using the volatility normalised moving average crossover divergence (MACD) indicators such that
\begin{align}
    & \text{MACD}(i, t, S, L) = m(i,t,S) - m(i, t, L) \\
    & \text{MACD}_{\text{norm}}(i, t, S, L) =  \frac{\text{MACD}(i, t, S, L)}{\text{std}(p_{i, t-63:t})} \\
    & y_{i,t}(S,L) =  \frac{\text{MACD}_{\text{norm}}(i, t, S, L) }{\text{std}(\text{MACD}_{\text{norm}}(i, t-252, S, L) )} \label{eq:MACD_Features} \\
    & x_{i,t} = \frac{1}{3} \sum_{k=1}^3 \phi( y_{i,t}(S_k,L_k)) \label{eq:MACD_position}
\end{align}%
where $(S_k, L_k) \in \{(8,24), (16, 48), (32, 96)\}$ and $\phi(y) = \frac{y\exp(-y^2/4)}{0.89}$ is a position scaling function.

\paragraph{Machine Learning-based TSMOM} Recent literature managed to construct TSMOM from machine learning models. \citet{limEnhancingTimeSeries2020} proposed a general supervised learning framework such that 
\begin{equation}
    x_{i,t} = \text{sgn}(y_{i,t}) \quad \text{and}\quad y_{i,t}=f_{\boldsymbol{\theta}}(\mathbf{u}_{i,t}). \label{eq:LinReg}
\end{equation}%
It can also directly model the position:
\begin{equation}
    x_{i,t} = f_{\boldsymbol{\theta}}(\mathbf{u}_{i,t})
\end{equation}%
The model specifications for $f$ are diverse and often considered in relation to the nature of the input feature $\mathbf{u}$. \citet{limEnhancingTimeSeries2020} compared linear models and deep learning models, such as Lasso Regression, MLP, WaveNet, and LSTM, using input features such as the MACD indicators and volatility-scaled returns from diverse lookback windows. In a similar vein, \citet{woodTradingMomentumTransformer2022} considered an attention-based deep learning model. \citet{tanSpatioTemporalMomentumJointly2023} and \citet{liuDeepInceptionNetworks2023} proposed advanced deep learning to effectively model for time-series and cross-sectional momentum. \par


\subsection{Network Momentum Strategy}

Network momentum strategy takes into account the interconnections of assets, assuming an asset's future performance can be influenced by the past performance of its linked assets. A network or graph is defined to represent economic links between assets, modelled by a graph adjacency matrix $\mathbf{A}_t$ at time $t$. A general framework of network momentum strategies starts by defining a \textit{network} or \textit{graph} to represent certain economic links between two assets. The network is mathematically modelled by a graph adjacency matrix $\mathbf{A}_t \in \mathbb{R}^{N_t \times N_t}$ between $N_t$ assets at time $t$, where the entries  $a_{ii, t} = 0$ and $a_{ij,t} \geq 0$. If $a_{ij,t} > 0$, there exists an economic link that shows certain similarity between asset $i$ and asset $j$. Some studies \cite{moskowitzIndustriesExplainMomentum1999, boniAnalystsIndustriesPrice2006} defined binary graph such that $a_{ij,t} \in \{0, 1\}$, while others \cite{yamamotoMomentumInformationPropagation2021, aliSharedAnalystCoverage2020} also quantified the strength of the interconnections using edge weights. \par

The network momentum of asset $i$ at time $t$ is defined through the average of trend estimation propagated from its connected assets in $\mathbf{A}_t$:
\begin{equation}
\label{eq: network_momentum}
     y_{i,t} = \frac{1}{|\mathcal{N}_t(i)|} \sum_{j \in \mathcal{N}_t(i)} a_{ij, t}  y_{j,t}  
\end{equation}%
where $\mathcal{N}_t(i)$ is the neighbourhood set of asset $i$ at time $t$ containing all its connected assets. In the literature, different economic links were explored. Some examples of $\mathbf{A}$ include:
\begin{itemize}

    \item \textit{Industry Linkage}. $a_{ij, t} = 1$ if stock $i$ and $j$ belong to the same industry, based on industrial classification codes such as two-digit SIC \cite{moskowitzIndustriesExplainMomentum1999} and GICS \cite{boniAnalystsIndustriesPrice2006}.


    \item \textit{Supply-demand Relations} \cite{menzlyMarketSegmentationCrosspredictability2010, yamamotoMomentumInformationPropagation2021}.  
    \citet{cohenEconomicLinksPredictable2008b} define a customer link $a_{ij, t} = 1$ for a target firm $i$ if it sells more then 10\% its products or services to the firm $j$. 
    
    \item \textit{Geographic Location}. $a_{ij, t} = 1$ if two companies have headquarters in the same location \cite{parsonsGeographicLeadLagEffects2020, korniotisStateLevelBusinessCycles2013a}.

    \item \textit{Patent Similarity}. \citet{leeTechnologicalLinksPredictable2019} define a technology proximity score $a_{ij,t}$ to measure the similarity of patent portfolios between two firms. 

    
    \item \textit{Analyst co-coverage}. \citet{aliSharedAnalystCoverage2020} defined $a_{ij, t}$ as the number of sell-side analysts who cover stock $i$ and $j$ in the same report at time $t$. 
\end{itemize}
While the above networks can effectively capture some economic similarity between companies and potentially generate alpha signals, they often rely on expensive databases from domain experts with deep understanding of finance. In this paper, we propose learning the graph adjacency matrix $\mathbf{A}_t$ from readily accessible historical returns that directly optimises momentum strategy. 

\subsection{Learning to Learn Graph Topologies}
\label{sec:preliminary:l2g}
Graph learning involves estimating a graph adjacency matrix by exploring the spectral characteristics presented in the node observations \cite{dongLearningGraphsData2019, kalofoliasHowLearnGraph2016, dongLearningLaplacianMatrix2016, mateosConnectingDotsIdentifying2019}. Given a feature matrix $\mathbf{V} \in \mathbb{R}^{N \times F}$ on $N$ nodes, the graph adjacency matrix can be optimised for by solving a convex optimisation problem such as the one in \cite{kalofoliasHowLearnGraph2016}: 
\begin{equation}
\label{eq: gl}
\begin{aligned}
    \min_{\mathbf{A}} & ~~  \text{tr}\Big(\mathbf{V}^{\top}(\mathbf{D} -\mathbf{A}) \mathbf{V} \Big)- \alpha \mathbf{1}^{\top} \log (\mathbf{A} \mathbf{1})+ \beta ||\mathbf{A}||_F^2  \\
    s.t. & ~~  \mathbf{A}_{ij} = \mathbf{A}_{ji}, ~~ \mathbf{A}_{ij} \geq 0 ~~ \forall i \neq j
\end{aligned}
\end{equation}%
where $\mathbf{D}$ is a diagonal matrix with $\mathbf{D}_{ii} = \sum_j \mathbf{A}_{ij}$. The first trace term measures the spectral variations of $\mathbf{V}$ on the learned graph, encouraging connections between nodes with similar features. The log and $\ell_2$ terms act as topological regularisation in the optimisation problem, enforcing graph connectivity to avoid isolated nodes and ensure a smooth edge weight distribution. The constraints ensure the learned graph adjacency matrix is symmetric and non-negative. \par


Although the optimisation problem can be efficiently solved by proximal gradient descent algorithms like primal-dual splitting (PDS), choices of hyperparameters $\alpha$ and $\beta$, significantly influence the resulting graph's topological structure, such as its sparsity. 
To address this, \citet{puLearningLearnGraph2021} introduced Learning to Learn Graph Topologies (L2G), a method reformulating the optimisation problem into a task of learning a parametric mapping. A neural network with a forward propagation step is proposed by unrolling the iterative PDS optimisation steps (as in Algorithm \ref{alg:unrolling}). The hyperparameters that influence the graph's topological characteristics transform into learnable parameters, which are trained using a loss function comparing graph estimates against ground truth graphs. \par


In our context, each financial asset represents a node, and their historical pricing data act as node observations. With graph learning, we can reveal the interconnections between assets. 
Learning to Learn Graph Topologies (L2G) provides further inspiration, as we can now use a loss function associated with portfolio performance to determine the optimal graph structure. 
\section{Proposed Methodology}

\subsection{L2GMOM}



In this section, we introduce \textit{Learning to Learn Financial Networks for Optimising Momentum Strategies} (L2GMOM), a solution specifically designed to learn financial networks from readily available data and optimise them directly for an ideal portfolio performance. 

As introduced in Section \ref{sec:preliminary:l2g}, L2G reformulates optimisation-based graph learning into an unrolling neural network. By leveraging the inherent modularity of neural networks, where different layers can be easily stacked for forward propagation, we propose to incorporate an additional layer into L2G for directly constructing network momentum. Given a feature matrix $\mathbf{V}_t$ captured from $N_t$ assets at time $t$, the initial step of L2GMOM involves employing the L2G layer (Algorithm \ref{alg:unrolling}), as defined in Eq.\eqref{eq:unrolling_layer}, to learn a graph adjacency matrix $\mathbf{A}_t$. This matrix illustrates the interconnections among the assets. Subsequently, $\mathbf{A}_t$ has a normalisation step, as outlined in Eq.\eqref{eq:g_norm_layer}. The normalisation aims to offset the dominance of nodes with many connections. In line with the machine learning-based TSMOM approach \cite{limEnhancingTimeSeries2020} and traditional network momentum in Eq.\eqref{eq: network_momentum}, we propose a linear layer to estimate the momentum trend of asset $i$, derived from a linear combination of individual momentum features $\mathbf{U}_t$ of its connected assets in $\tilde{\mathbf{A}}_t$. $\tilde{\mathbf{A}}_t \mathbf{U}_t$ has an interpretation of a network momentum features and the learned parameters $\boldsymbol{\theta}$ are the weights of combining different momentum features. It is worth mentioning that $\tilde{\mathbf{A}}_t$ from Eq.\eqref{eq:g_norm_layer} contains only non-negative edges, as a result of enforcing the constraints of Eq.\eqref{eq: gl}.   
In summary, these are the forward propagation rules for the L2GMOM neural network:
\begin{subequations}
\begin{align}
&  \mathbf{A}_t = \textbf{L2G}(\mathbf{V}_t, L; ~ \boldsymbol{\alpha}, \boldsymbol{\beta}, \boldsymbol{\gamma}) \label{eq:unrolling_layer} \\
& \tilde{\mathbf{A}}_t = \mathbf{D}_t^{-1/2}\mathbf{A}_t \mathbf{D}_t^{-1/2}, ~\text{where} ~\mathbf{D}_{ii,t} = \sum_j \mathbf{A}_{ij,t}, ~\mathbf{D}_{ij,t} = 0 \label{eq:g_norm_layer} \\
& (\textbf{L2GMOM}) \quad \mathbf{y}_{t}= \tilde{\mathbf{A}}_t \mathbf{U}_t \boldsymbol{\theta} + b \label{eq:l2gmom}
\end{align}
\end{subequations}%
where $\mathbf{y}_{t}$ is the model output, a $N_t$-dimensional vector of trend estimations for every assets, such that $\mathbf{y}_{t} = [y_{1,t}, y_{2,t}, \dots, y_{N_t,t}]^{T}$.
We further take the trading positions $\mathbf{x}_{t} =  \text{sgn}(\mathbf{y}_{t})$ for constructing a network momentum strategy. The trainable parameters include $(\boldsymbol{\alpha}, \boldsymbol{\beta}, \boldsymbol{\gamma}, \boldsymbol{\theta}, b)$.

\begin{algorithm}[]
    \caption{\textbf{L2G} (PDS unrolling)}
    \label{alg:unrolling}
     \begin{algorithmic}[1]
     \renewcommand{\algorithmicrequire}{\textbf{Input:}}
     \renewcommand{\algorithmicensure}{\textbf{Output:}}
     \REQUIRE Feature matrix $\mathbf{V} \in \mathbb{R}^{N \times F}$, number of unrolling times $L$ 
     \STATE Initial primal $\mathbf{w}_0 = \mathbf{0} \in \mathbb{R}^{N(N-1)/2}$ and dual $\mathbf{v}_0 = \mathbf{0} \in \mathbb{R}^{N}$ \par
     \STATE $\mathbf{h} = \text{vech}(\mathbf{H})$, where $\mathbf{H}_{ij} = ||\mathbf{V}_{i,:} - \mathbf{V}_{j,:}||^2_2$
     \STATE Degree operator $\mathbf{D}$ such that $\mathbf{Dw} = [d_1, \dots, d_N]^T$, where $d_i = \sum_j {\mathbf{W}_{ij}}$ and $\mathbf{w} = \text{vech}(\mathbf{W})$
     \FOR {$l = 0, 1, \dots, L$}
       \STATE  $\mathbf{r}_{1, l} = \mathbf{w}_l - \gamma_l (2 \beta_l \mathbf{w}_l + 2\mathbf{h} + \mathbf{D}^T \mathbf{v}_l)$
       \STATE  $\mathbf{r}_{2,l} = \mathbf{v}_{l} +  \gamma_l \mathbf{D} \mathbf{w}_l$
      
       
       \STATE $\mathbf{p}_{1,l} = \max\{\mathbf{0}, \mathbf{r}_{1,l}\}$
      
       \STATE $\mathbf{p}_{2,l}  = \big( \mathbf{r}_{2,l} - \sqrt{\mathbf{r}_{2,l}^2 + 4 \alpha_l \gamma_l} \big)/2$
      
       \STATE $\mathbf{q}_{1,l} = \mathbf{p}_{1,l} -  \gamma_l (2 \beta_l \mathbf{p}_{1,l} + 2\mathbf{h} + \mathbf{D}^T \mathbf{p}_{2,l})$
       
       \STATE $\mathbf{q}_{2,l} = \mathbf{p}_{2,l} + \gamma_l \mathbf{D} \mathbf{p}_{1,l}$
    
       \STATE $\mathbf{w}_{l+1} = \mathbf{w}_l - \mathbf{r}_{1,l} + \mathbf{q}_{1,l}$
       
       \STATE $\mathbf{v}_{l+1} = \mathbf{v}_{l} - \mathbf{r}_{2,l} + \mathbf{q}_{2,l}$
    \ENDFOR
    \RETURN Graph estimate $\mathbf{A}$ such that $\mathbf{w}_L = \text{vech}(\mathbf{A})$
    \end{algorithmic} 
\end{algorithm}




Taking one step further, we also model the direct position $\mathbf{x}_t$ instead of the trend estimation $\mathbf{y}_t$. We name the model \textbf{L2GMOM\_SR}, since we further use Sharpe Ratio (SR) to optimise for the positions, which are consistent with the literature \cite{limEnhancingTimeSeries2020}. Since the resulting positions $\mathbf{x}_t$ lie within the range of $[-1, 1]$ in TSMOM, we simply introduce a non-linear activation function, tanh, in the linear layer to ensure the same bahaviour:
\begin{equation}
(\textbf{L2GMOM\_SR})  \quad \mathbf{x}_t  = \text{tanh}\Big( \tilde{\mathbf{A}}_t\mathbf{V}_t \boldsymbol{\theta} + b  \Big) \label{eq:l2gmom_sr}
\end{equation}%
The use of L2GMOM\_SR showcases our model's flexibility, accommodating different loss functions. L2GMOM\_SR not only optimises the portfolio positions, but also the graph learning layer, ensuring the network is well optimised w.r.t. Sharpe ratio. 

\subsection{Input Features} \label{sec:features}
We consider the following eight momentum features calculated from pricing data for $\textbf{U}_{t}$:
\begin{itemize}
    \item{\textbf{Volatility normalised returns}} over the past 1-day, 1-month, 3-month, 6-month and 1-year periods; defined as $\rfrac{r_{t-\Delta:t}}{\sigma_t \sqrt{\Delta}}$, where $\sigma_t$ is an ex-ante daily volatility estimated from an exponential weighted moving standard deviation with 60-day span on daily return. 
    \item{\textbf{MACD} indicators} with three combinations of short and long time scales $(S_k, L_k) \in \{(8,24), (16, 48), (32, 96)\}$ in Eq.\eqref{eq:MACD_Features}.
\end{itemize}%
Therefore, $\mathbf{U}_t \in \mathbb{R}^{N_t\times 8}$ is the matrix containing these eight features for every asset. 

For $\textbf{V}_{t}$, we adopt a 252-day lookback window and stack the individual features together. This is to ensure the L2G layer can capture a stable similarity from longer time-series, providing a robust foundation for learning. As a result, we obtain a representation $\mathbf{V}_t \in \mathbb{R}^{N_t\times 8\delta}$, where $\delta = 252$. \par

In order to mitigate the influence of outliers, we also winsorise the data by capping/flooring it to be within 5 times its exponentially weighted moving (EWM) standard deviations from its EWM average – computed using a 252-day half-life, as adopted by \citet{limEnhancingTimeSeries2020}.


\subsection{Loss Functions}

The parameters in L2GMOM to be estimated include graph learning parameters $(\boldsymbol{\alpha}, \boldsymbol{\beta}, \boldsymbol{\gamma})$ and momentum feature parameters $(\boldsymbol{\theta}, b)$. Note that \citet{puLearningLearnGraph2021} observed that allowing varying hyperparameters $(\alpha, \beta, \gamma)$ in each unrolling layer would significantly increase the performance of graph learning. Therefore, we adopt the same approach. The coefficients are learned by minimising with respect to different loss functions associated with portfolio performance, depending on whether the output is $\mathbf{y}_t$ or $\mathbf{x}_t$.
\begin{itemize}
    \item{Mean Squared Error (MSE)} for training L2GMOM:
    \begin{equation} \label{eq:loss_MSE}
        \mathcal{L}_{\text{MSE}}(\boldsymbol{\theta}, b) = \frac{1}{|\Omega|}\sum_{(i,t) \in \Omega}\left( y_{i,t} - \frac{r_{i, t:t+1}}{\sigma_{i,t}} \right)^2
    \end{equation}

    \item{Negative Sharpe Ratio} for training L2GMOM\_SR:
    \begin{equation} \label{eq:loss_SR}
        \mathcal{L}_{\text{SR}}(\boldsymbol{\theta}, b) = - \frac{\sum_{(i,t) \in \Omega} R_{i,t} \times \sqrt{252}}{\sum_{(i,t) \in \Omega} R_{i,t}^2 - (\sum_{(i,t) \in \Omega} R_{i,t})^2}
    \end{equation}%
    where $R_{i,t} = x_{i,t} \frac{\sigma_{\text{tgt}}}{\sigma_{i,t}} r_{i, t:t+1}$ is the volatility-normalised return of asset $i$ at time $t$.
\end{itemize}

\section{Performance Evaluation}

\label{sec:backtest}

\subsection{Backtest Setup}
\subsubsection{Datasets}
Our dataset contains 64 ratio-adjusted continuous future contracts obtained from the Pinnacle Data Crop CLC Database\footnote{\href{https://pinnacledata2.com/clc.html}{https://pinnacledata2.com/clc.html}}. We computed the momentum feature in Section \ref{sec:features} from the daily prices from 1990 to 2020, although different assets could have different time spans when the prices are available. In Appendix, Table \ref{table:universe} lists all the assets used in backtest. 
 
\subsubsection{Strategy Candidates}
We compare the proposed strategies against the following reference benchmarks:
\begin{itemize}
    \item{\textbf{Long Only}} takes a consistent long position $x_{i,t} = 1$ with daily volatility scaling in Eq.\eqref{eq: TSMOM}
    \item{\textbf{MACD}} \cite{bazDissectingInvestmentStrategies2015} takes positions from Eq.\eqref{eq:MACD_position}.
    \item{\textbf{LinReg}} \cite{limEnhancingTimeSeries2020} stands for Linear Regression with the eight \textit{momentum features} as covariates, and the signals were constructed following Eq.\eqref{eq:LinReg} . 
    \item{\textbf{GLinReg}} stands for Graph Linear Regression. We learn graphs separately with Eq.\eqref{eq: gl} and we fit them in Eq.\eqref{eq:l2gmom}. This is to demonstrate whether the performance gain is from the end-to-end learning architecture. 
\end{itemize}
We name the strategies constructed from the proposed models as \textbf{L2GMOM} and \textbf{L2GMOM\_SR}. The positions of L2GMOM are the signs of the trend estimation in Eq.\eqref{eq:l2gmom}, and L2GMOM\_SR directly outputs the positions.

\subsubsection{Optimisation Details} 
The models were trained every 5 years, involving coefficient optimisation and hyperparameter tuning using data up to the re-trained time point. These optimised models were then used to generate trading signals for the subsequent 5 years, using out-of-sample data. To ensure an adequate number of training samples, the first backtest period covered 1990-1999 for training and 2000-2004 for testing. Similarly, the second period included 1990-2004 for training and 2005-2009 for testing, and so on. Thus, we had a 20-year aggregated out-of-sample period from 2000 to 2020. For each backtest period, the most recent 10\% of training data was set aside as a validation set for hyperparameter tuning. The hyperparameters included the number of unrolling times $L$ in Algorithm \ref{alg:unrolling}, batch size, and optimiser parameters. Linear models such as LinReg and GLinReg have analytical solutions. For L2GMOM and L2GMOM\_SR, we used stochastic gradient descent with the \texttt{Adam} optimiser in a mini-batch approach. Early stopping was activated when the validation loss failed to improve beyond a certain threshold, which was determined using the validation set for convergence. To mitigate the randomness introduced by the \texttt{Adam} optimiser, we conducted 5 runs with random initial values and averaged the model outputs to create an ensemble. \par

\subsection{Portfolio Performance}

\begin{table*}[t]
\caption{Portfolio Performance Metrics}
\label{tab:bt_perf_table_combined}

\begin{threeparttable}
\begin{tabular}{lllllllllll}
\toprule
{} &    return &    vol. &  Sharpe &  \makecell{downside \\ deviation} & \makecell{MDD} &  \makecell{MDD \\ duration} &  Sortino &  Calmar &  hit rate &  $\frac{\text{Avg. P}}{\text{Avg. L}}$ \\
\midrule
\multicolumn{11}{l}{\textbf{Raw Signals:}  } \\
Long Only & 0.036 & 0.053 &   0.671 &               0.037 &         0.200 &                 17.2\% &    0.951 &   0.173 &    53.9\% &                 0.955 \\
MACD      & 0.025 & 0.047 &   0.533 &               0.033 &         0.121 &                 27.5\% &    0.766 &   0.200 &    52.9\% &                 0.977 \\
LinReg    & 0.043 & 0.042 &   1.029 &               0.028 &         0.080 &                 11.5\% &    1.541 &   0.538 &    53.6\% &                 1.031 \\
GLinReg   & 0.060 & 0.049 &   1.227 &               0.033 &         0.085 &                  7.7\% &    1.806 &   0.713 &    \bf{54.3\%} &                 1.034 \\
L2GMOM    & \bf{0.065} & 0.047 &   \bf{1.400} &               0.030 &         0.066 &                  7.2\% &    \bf{2.197} &   \bf{0.942} &    50.7\% &                 \bf{1.086} \\
L2GMOM\_SR & 0.045 & \bf{0.032} &   1.394 &               \bf{0.023} &         \bf{0.052} &                  \bf{6.2\%} &    1.911 &   0.817 &    52.0\% &                 1.051 \\
\midrule
\multicolumn{11}{l}{\textbf{Signals Rescaled to 15\% Target Volatility:}  } \\
Long Only & 0.117 & 0.150 &   0.792 &               0.095 &         0.633 &                 17.1\% &    1.229 &   0.177 &    53.9\% &                 0.969 \\
MACD      & 0.102 & 0.150 &   0.694 &               0.094 &         0.342 &                 23.7\% &    1.085 &   0.279 &    52.9\% &                 0.997 \\
LinReg    & 0.174 & 0.150 &   1.189 &               0.092 &         0.285 &                 11.9\% &    1.892 &   0.623 &    53.6\% &                 1.049 \\
GLinReg   & 0.200 & 0.150 &   1.365 &               0.093 &         0.253 &                  8.9\% &    2.149 &   0.822 &    \bf{54.3\%} &                 1.048 \\
L2GMOM    & 0.224 & 0.150 &   1.525 &               \bf{0.089} &         0.186 &                  7.2\% &    2.531 &   1.201 &    50.7\% &                 \bf{1.099} \\
L2GMOM\_SR & \bf{0.255} & 0.150 &   \bf{1.744} &               0.094 &         \bf{0.173} &                  \bf{5.7\%} &    \bf{2.716} &   \bf{1.495} &    52.0\% &                 1.091 \\
\bottomrule
\end{tabular}
    \begin{tablenotes}
      \item[a] Best performance of each metric is \bf{bold}.
    \end{tablenotes}
\end{threeparttable}
\end{table*}

\begin{figure*}[h]
    \centering
    \begin{subfigure}[]{0.49\textwidth}
        \includegraphics[width=1\textwidth]{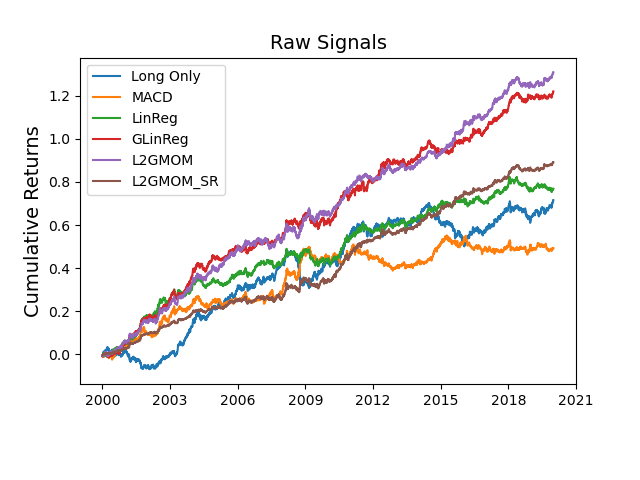}
        \caption{Raw signals}
        \label{subfig:cumulative_returns_raw}
    \end{subfigure}
    \hspace{0.1cm}
    \begin{subfigure}[]{0.49\textwidth}
        \includegraphics[width=1\textwidth]{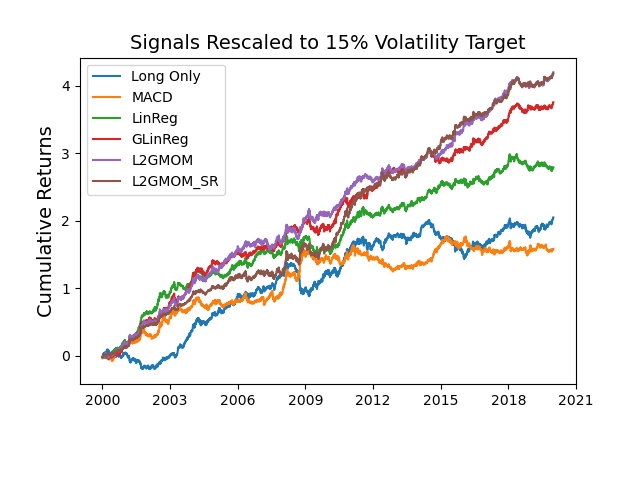}
        \caption{Signals rescaled to 15\% volatility target}
        \label{subfig:cumulative_returns_scaled}
    \end{subfigure}
\caption{The cumulative daily returns of the proposed strategy (L2GMOM/L2GMOM\_SR) and four reference strategies (Long Only/MACD/LinReg/GLinReg) in different colours for the entire out-of-sample period from 2000 to 2020.}
\label{fig:cumulative_returns}
\end{figure*}

In evaluating the portfolio performance, we use the following annualised metrics in Table \ref{tab:bt_perf_table_combined}:
\begin{itemize}
    \item{\textbf{Profitability}}: expected return and hit rate (the percentage of days with positive returns across the test period)
    \item{\textbf{Risk}}: daily volatility (vol.), downside deviation, maximum drawdown (MDD) and MDD duration (the percentage of days with MDD across the test period).
    \item{\textbf{Overall Performance}}: Sharpe ratio (expected return$/$vol.), Sortino ratio (expected return$/$downside deviation), Calmar ratio (expected return$/$MDD), and the average profits over the average loss $\left( \frac{\text{Avg. P}}{\text{Avg. L}} \right)$.
\end{itemize}%
In order to have a better comparison between different strategies, we apply an additional layer of volatility scaling at the portfolio level to an annualised volatility target of 15\% and report the above evaluation metrics. In Figure \ref{subfig:cumulative_returns_raw} and Figure \ref{subfig:cumulative_returns_scaled}, we plot the daily cumulative returns of raw signals and signals rescaled to 15\% annual volatility of main strategies. 

L2GMOM and GLinReg both outperform LinReg in terms of portfolio returns in both raw signals and signals rescaled to 15\% target volatility. This suggests that considering the network effect in assets can greatly enhance profitability. L2GMOM also outperforming GLinReg indicates that optimising the graph with respect to portfolio performance can yield higher returns, which supports our conjecture that the separation of network construction in network momentum strategies may adversely affect the portfolio performance. In terms of portfolio risk, L2GMOM has a lower downside deviation and shorter MDD duration in both raw signals and signals rescaled to 15\% target volatility, indicating that it manages risk more effectively. L2GMOM also outperforms GLinReg in terms of volatility, suggesting that optimising the graph with respect to portfolio performance can also reduce risk. \par

L2GMOM\_SR has the lowest volatility, downside deviation, MDD, and MDD duration in the raw signal, indicating that optimising with the Sharpe ratio can significantly reduce risk. It is not surprising to observe a higher Sharpe ratio in L2GMOM\_SR, which is consistent with the results in machine learning TSMOM \cite{limEnhancingTimeSeries2020}. Note that, the returns of L2GMOM\_SR drops in the raw signals, but the return is the highest in the volatility-scaled signals. \par

\subsection{Diversification Analysis}

\begin{figure}[h!]
    \centering
    \begin{subfigure}[]{0.25\textwidth}
        \includegraphics[width=1\textwidth]{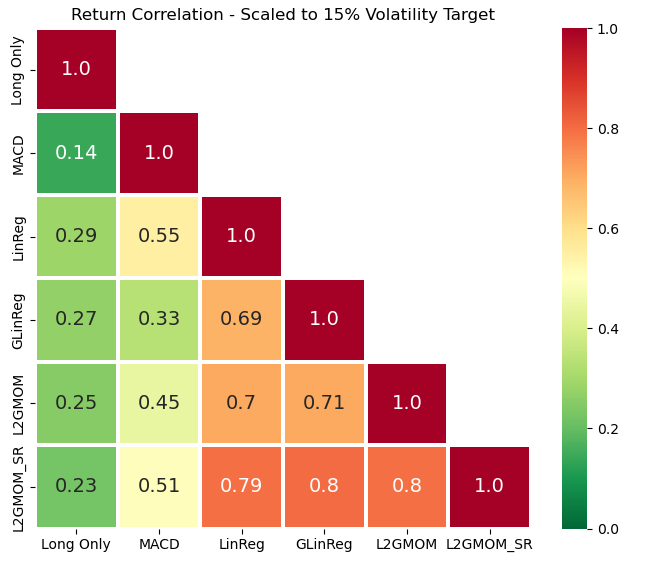}
        \caption{Return Correlations}
        \label{subfig:bt_corr}
    \end{subfigure}%
    \begin{subfigure}[]{0.25\textwidth}
        \includegraphics[width=1\textwidth]{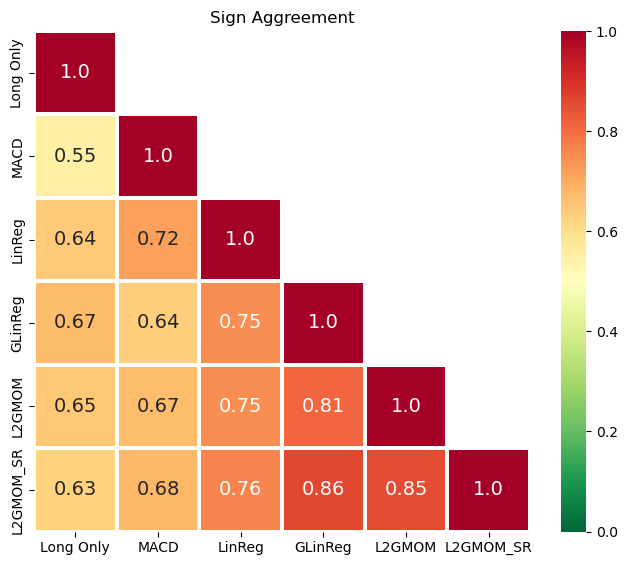}
        \caption{Sign Agreement}
        \label{subfig:sign_agreemtn}
    \end{subfigure}
\caption{A diversification analysis on the correlations and sign agreement of daily returns between the proposed strategies (L2GMOM/L2GMOM\_SR) and four reference strategies (Long Only/MACD/LinReg/GLinReg) of the entire out-of-sample period from 2000 to 2020. The values of the pairwise correlation and sign agreement are marked in the corresponding square boxes. 
}
\label{fig:diversification}
\end{figure}

To evaluate whether the signals constructed from the proposed models are different from the reference baselines, we calculated the \textit{correlation} of daily returns between candidate signals in Figure \ref{subfig:bt_corr}. Besides, we calculate the percentage by how much the signals of two strategies have the same direction, also known as \textit{sign agreement} in Figure \ref{subfig:sign_agreemtn}. Note that LineReg, GLinReg, L2GMOM and L2GMOM\_SR have the exactly same input momentum features. \par

From Figure \ref{subfig:bt_corr} and Figure \ref{subfig:sign_agreemtn}, we draw the following conclusions. The L2GMOM strategy, which incorporates graph learning and network momentum, shows moderate diversification potential with the model-free MACD strategy, as evidenced by their moderate return correlation and position sign agreement. Comparing L2GMOM with LinReg, which uses the same input but different momentum calculation methods, shows less diversification potential due to higher correlation and agreement. However, L2GMOM has better portfolio performance, showing it could be a better alternative strategy. Finally, L2GMOM, GLinReg, and L2GMOM\_SR, all of which consider graph effects but differ in optimisation methods, show limited diversification potential due to high return correlations and sign agreements. Despite their different modelling approaches, the similar inputs and graph-based methods used by these strategies result in relatively high correlations, suggesting limited diversification benefits when combined in a portfolio.

\subsection{Turnover Analysis}
Following the convention of literature \cite{limEnhancingTimeSeries2020}, we calculate the cost-adjusted Sharpe ratio from the adjusted returns:
\begin{equation}
    \tilde{r}_{i, t:t+1}^{\text{portfolio}} := \frac{1}{N_t}\sum_{i=1}^{N_t}\Big( x_{i,t}\frac{\sigma_{\text{tgt}}}{\sigma_{i,t}}r_{i, t:t+1} - c \cdot \sigma_{i,t} \Big| \frac{x_{i,t}}{\sigma_{i,t}} - \frac{x_{i,t-1}}{\sigma_{i,t-1}} \Big| \Big)
\end{equation}%
where $c$ is the turnover cost in bps, varying from 0.5 to 5 bps. As ex-ante costs rise, all strategies see a drop in cost-adjusted Sharpe ratios. LinReg, a machine learning model, is most sensitive, turning negative at costs over 3 bps. GLinReg, which includes graph learning, fares better, maintaining a higher positive Sharpe ratios at 2 bps. The proposed models, L2GMOM and L2GMOM\_SR, also stay positive at 3 bps, showing resilience to increasing costs. Thus, graph learning strategies like GLinReg, L2GMOM, and L2GMOM\_SR offer better cost resilience than LinReg. The Long Only and MACD strategies, which are model-free, show a relatively gradual decrease in Sharpe ratio as ex-ante costs increase. To further reduce the turnover of machine learning strategies, \cite{limEnhancingTimeSeries2020} proposed including a turnover regularisation. This is an interesting avenue for future analysis.

\begin{figure}[h!]
\centering
\includegraphics[width=0.45\textwidth]{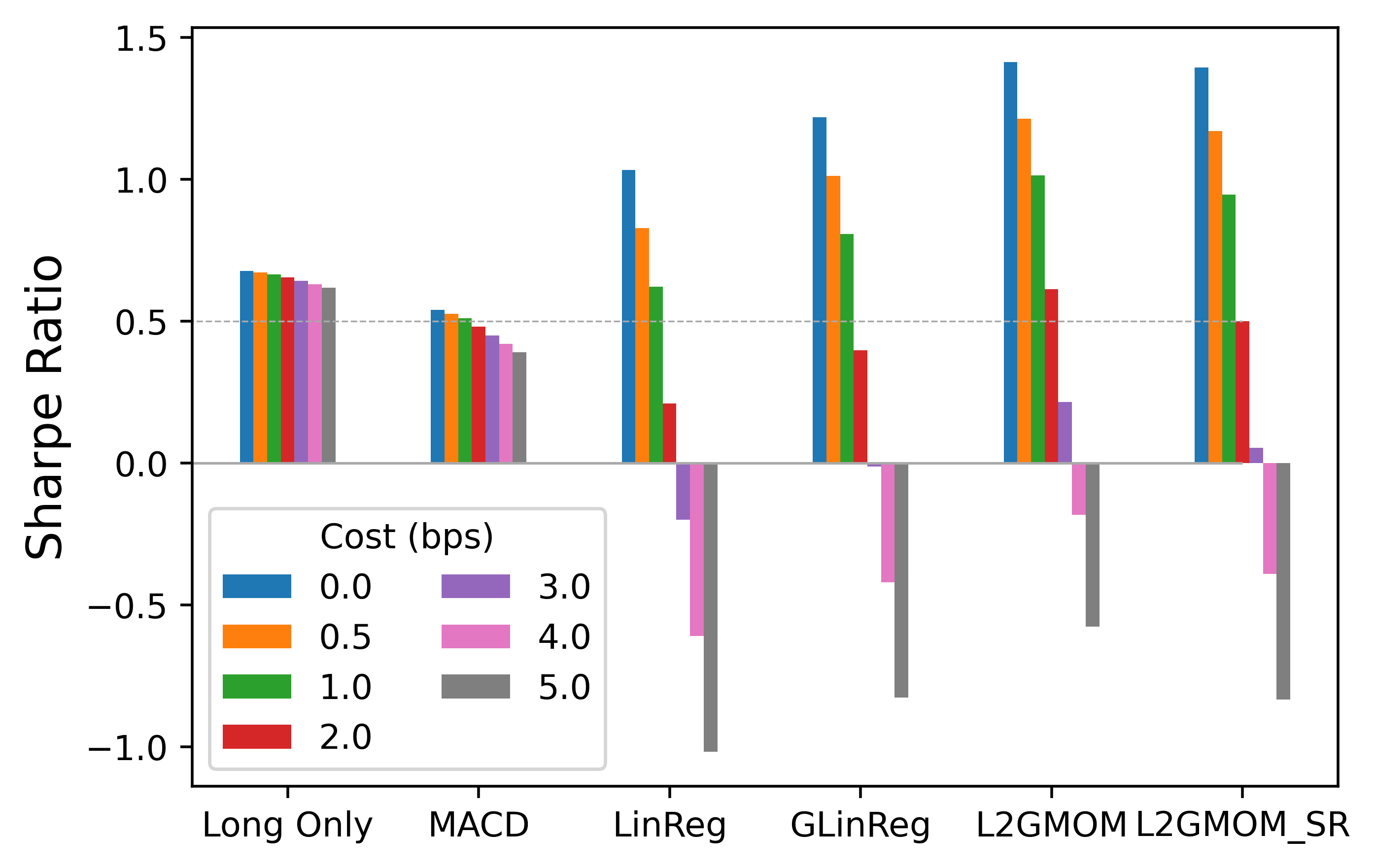}
\caption{The cost-adjusted Sharpe ratio of the proposed strategies (L2GMOM/L2GMOM\_SR) and four reference strategies (Long Only/MACD/LinReg/GLinReg) of the entire out-of-sample period from 2000 to 2020.}
\label{fig:sharpe-turnover}
\end{figure}

\subsection{Network Analysis}


\begin{figure*}[]
\centering
    \begin{subfigure}[]{0.49\textwidth}
        \includegraphics[width=0.9\textwidth]{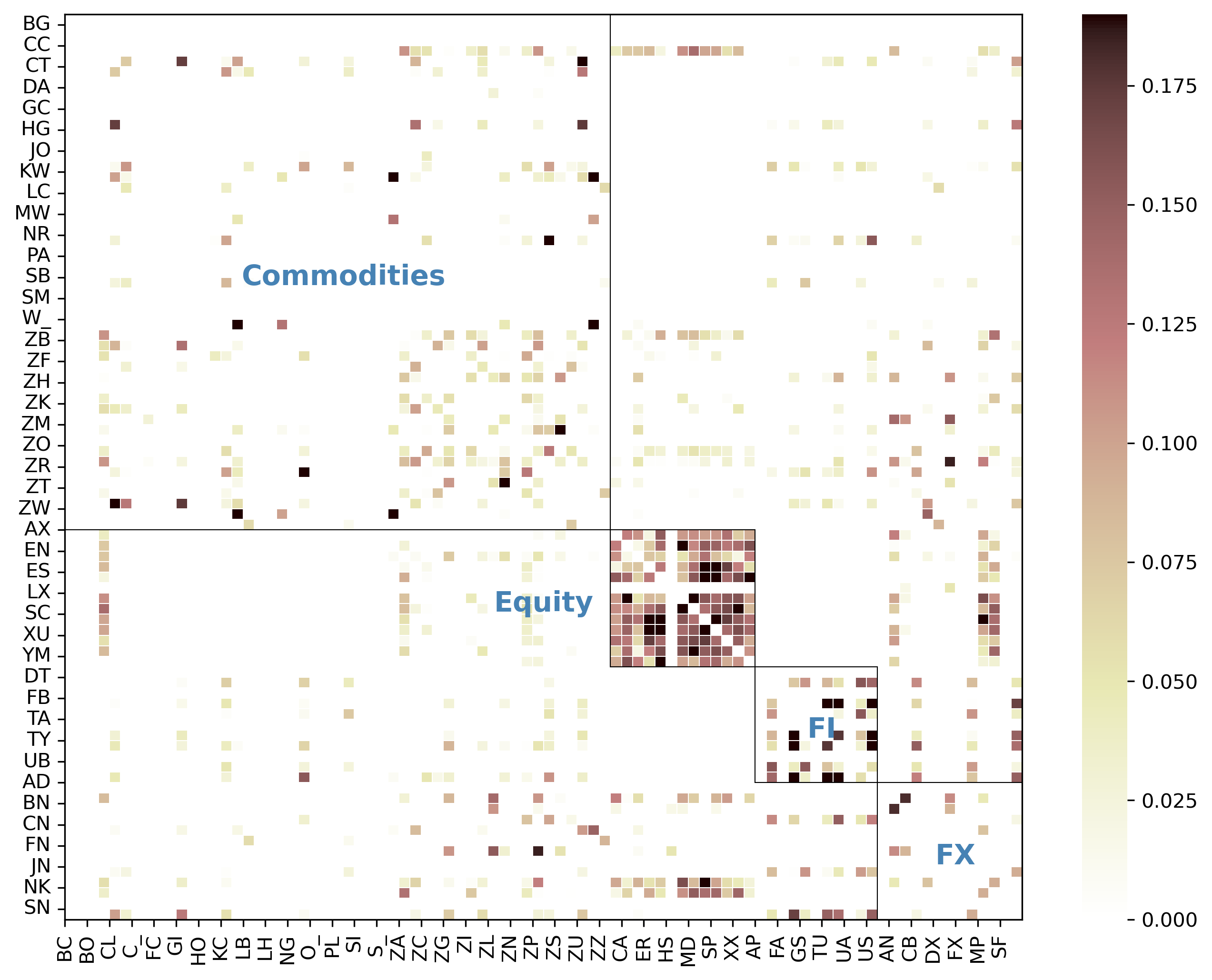}
        \caption{2008-01-03 from L2GMOM}
        \label{subfig:graph_l2gmom_2008}
    \end{subfigure}
    \hspace{0.1cm}
    \begin{subfigure}[]{0.49\textwidth}
        \includegraphics[width=0.9\textwidth]{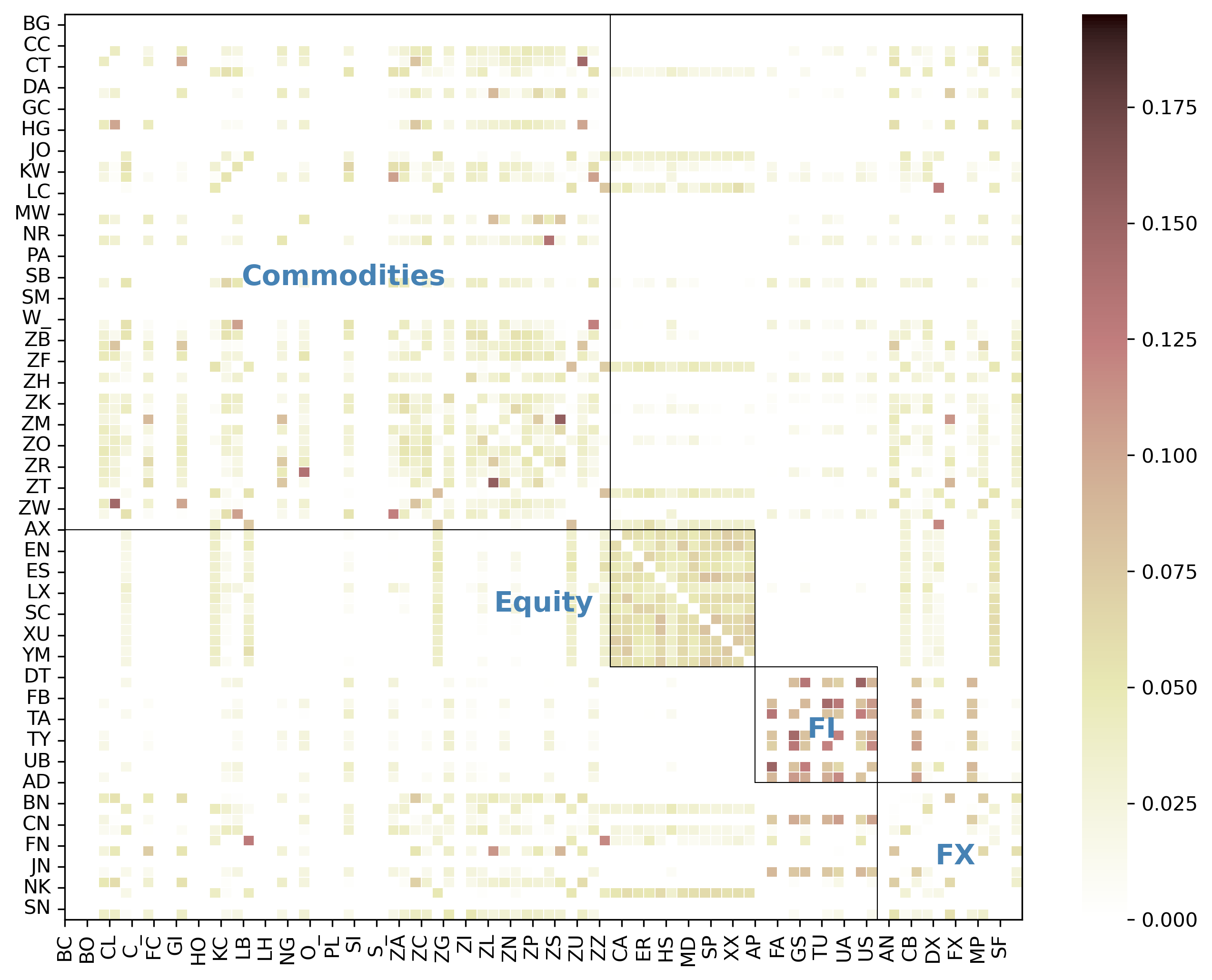}
        \caption{2008-01-03 from L2GMOM\_SR}
        \label{subfig:graph_l2gmom_sr_2008}
    \end{subfigure}
    \begin{subfigure}[]{0.49\textwidth}
        \includegraphics[width=0.9\textwidth]{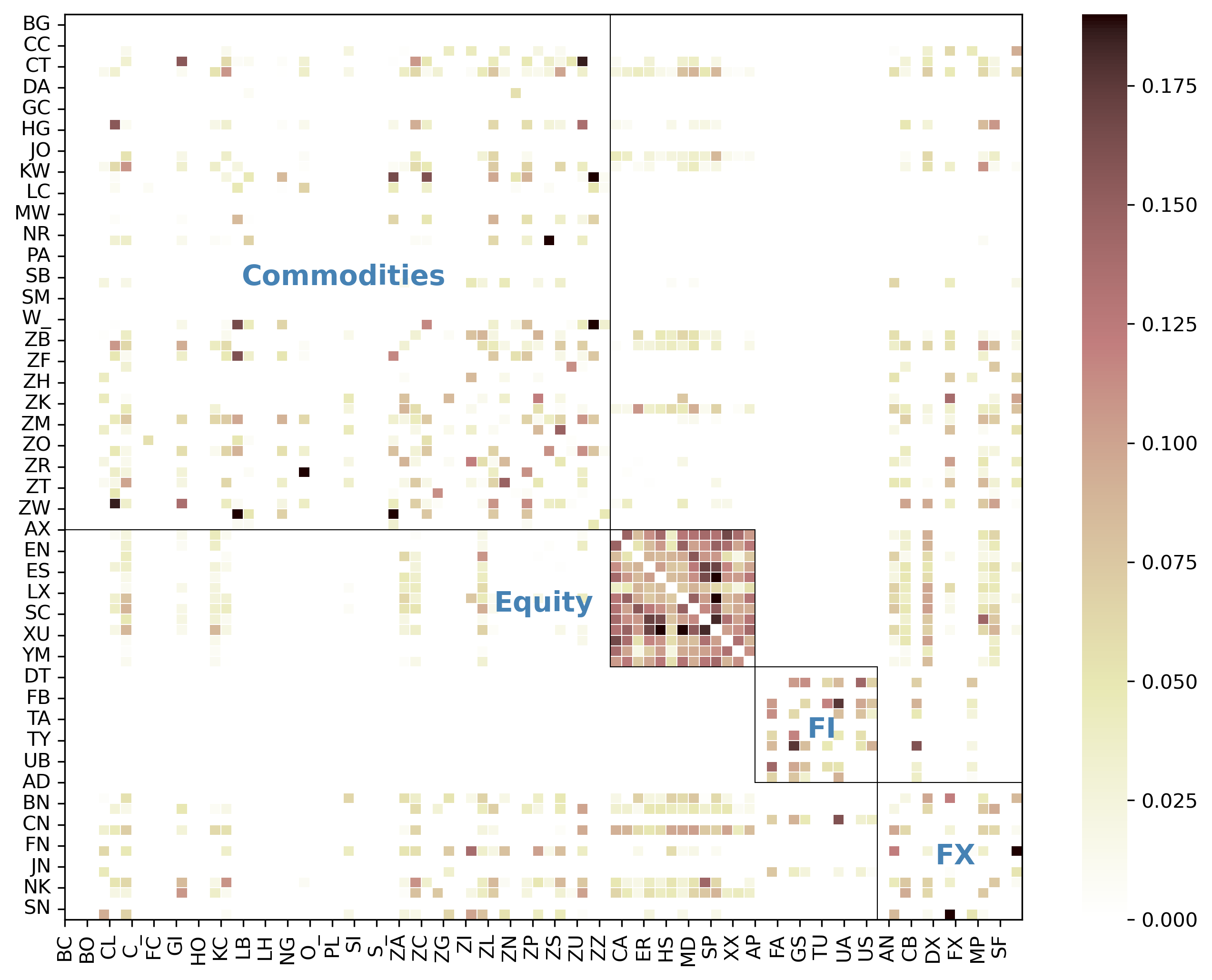}
        \caption{2010-01-04 from L2GMOM}
        \label{subfig:graph_l2gmom_2010}
    \end{subfigure}
    \hspace{0.1cm}
    \begin{subfigure}[]{0.49\textwidth}
        \includegraphics[width=0.9\textwidth]{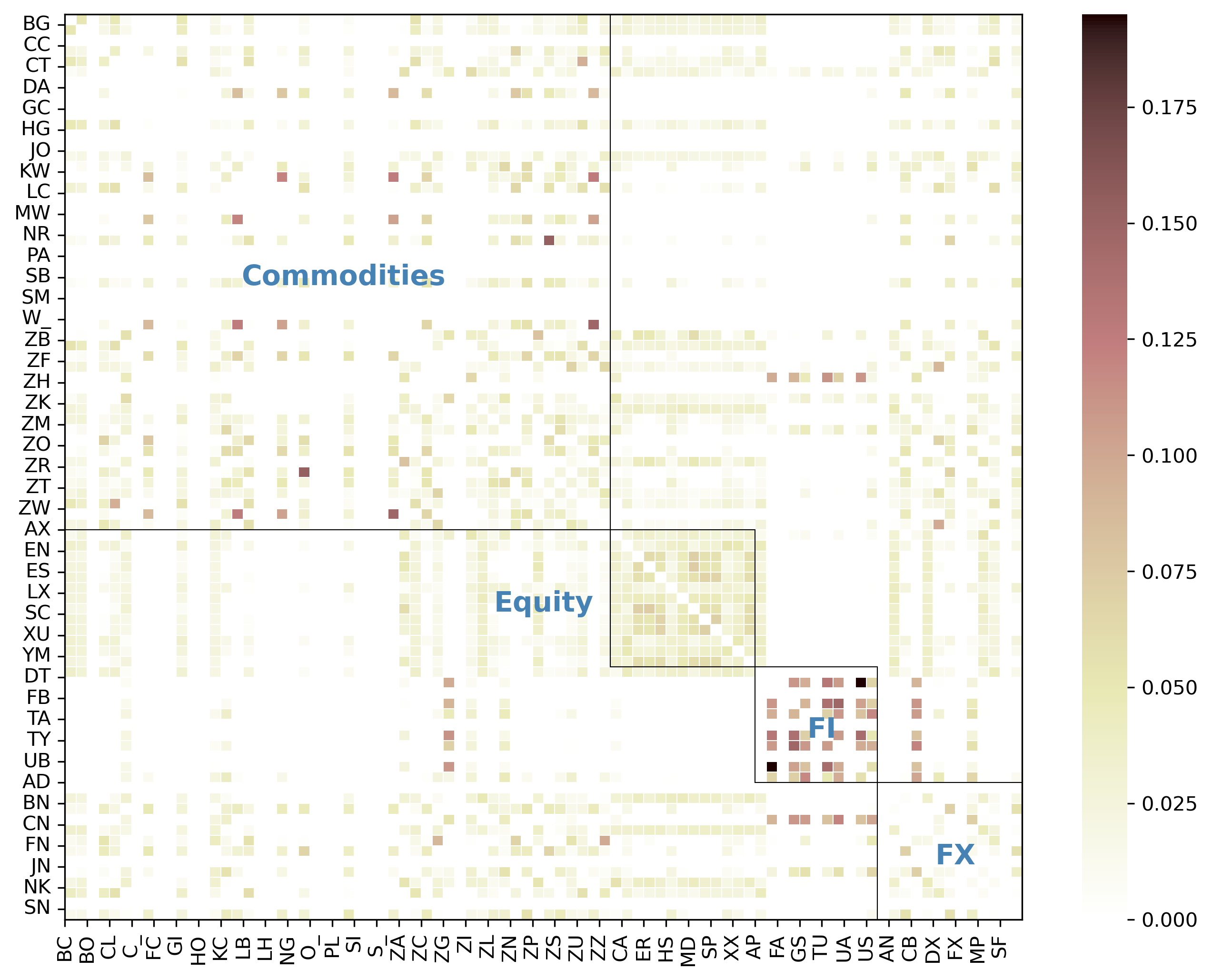}
        \caption{2010-01-04 from L2GMOM\_SR}
        \label{subfig:graph_l2gmom_sr_2010}
    \end{subfigure}
\caption{Momentum Spillover Analysis. Networks obtained from L2GMOM or L2GMOM\_SR. }
\label{fig:graphs}
\end{figure*}

Figure \ref{fig:graphs} are example daily networks obtained from the proposed L2GMOM and L2GMOM\_SR, specifically the output of Eq.\eqref{eq:g_norm_layer}. We deliberately chose two different dates to exhibit, 2008-01-03 for the turbulent period and 2010-01-04 for the calm period. Note that there were some assets without any connections, because the input features for them were missing. The four subfigures demonstrate a strong community structure corresponding to four asset classes: commodities, equities, fixed income (FI), and foreign exchange (FX). The equities form a dense cluster, indicating a high degree of interconnection among equity assets. Inter-class connections are also observed, particularly between commodities and the other three asset classes. However, there is a notable absence of connections between equities and fixed income. This is reasonable given that these two asset classes often exhibit negative correlations and are used to hedge each other, suggesting no momentum spillover according to the strategy. \par

By comparing Figure \ref{subfig:graph_l2gmom_2008} and \ref{subfig:graph_l2gmom_2010}, we can also capture some temporal variations in the momentum network structure. During calm periods (e.g. 2010-01-04), there are fewer inter-class connections between fixed income and the other assets. However, in turbulent times (e.g. Global Financial Crisis covering 2008-01-03), these connections increase, indicating a shift in momentum propagation under different market conditions. \par

The graphs obtained from the L2GMOM\_SR strategy are denser with more similar edge weights, demonstrating the strategy's tendency in reducing volatility. One possible explanation is that, by increasing the connections between assets, the strategy ensures that each asset receives momentum information passed from a broader range of assets, rather than relying on a few, thus enhancing the robustness and stability of the strategy.

\section{Conclusion}

In this paper, we present an innovative end-to-end graph machine learning framework for the construction of network momentum strategies, specifically two model variants: L2GMOM and L2GMOM\_SR. Our framework effectively learns the graph adjacency matrix and optimises the momentum strategy directly from historical pricing data, reducing the dependence on costly linkage databases and financial expertise typically needed for constructing network momentum factors, such as those derived from supply chain data.

Our models exceed the performance of traditional time series momentum strategies in both portfolio returns and risk management across 64 continuous futures contracts. Moreover, our models are highly interpretable, providing explicit networks as a byproduct. These networks not only represent the similarity between the momentum features of these assets, but are also optimised with respect to portfolio performance. This offers a powerful interpretation of the interconnections that can aid in constructing network momentum strategies. By analysing these learned networks, we uncover the community structure and temporal variations that align with common financial sense.



\bibliographystyle{ACM-Reference-Format}
\bibliography{0bib}


\appendix

\section{Dataset Details}
\label{sec:appendix-dataset}

\begin{table*}[b]
\centering
\footnotesize
\caption{The Pinnacle Universe}
\label{table:universe}
\begin{tabular}{lll|lll}
\toprule
\textbf{Ticker} & \textbf{Description} & \textbf{Period} & \textbf{Ticker} & \textbf{Description} & \textbf{Period}\\
\midrule
\multicolumn{2}{l}{\textbf{Commodities:}  } & & \multicolumn{2}{l}{\textbf{Equities:}  } \\
BC & BRENT CRUDE OIL, composite & 2010-2020 & AX & GERMAN DAX INDEX & 1999-2020 \\
BG & BRENT GASOIL, composite & 2010-2020 & CA & CAC40 INDEX & 2000-2020 \\
CC & COCOA & 1990-2020 & EN & NASDAQ, MINI & 2001-2020 \\
CL & CRUDE OIL & 1990-2020 & ER & RUSSELL 2000, MINI & 2004-2020 \\
CT & COTTON \#2 & 1990-2020 & ES & S\&P 500, MINI & 1999-2020 \\
DA & MILK III, composite & 1999-2020 & HS & HANG SENG INDEX & 1999-2020 \\
GI & GOLDMAN SAKS C. I. & 1995-2020 & LX & FTSE 100 INDEX & 1991-2020\\
JO & ORANGE JUICE & 1990-2020 & MD & S\&P 400, MINI, electronic & 1994-2020 \\
KC & COFFEE & 1990-2020 & SC & S\&P 500, composite & 1996-2020 \\
KW & WHEAT & 1990-2020 & SP & S\&P 500, day session & 1990-2020 \\
LB & LUMBER & 1990-2020 & XU & DOW JONES EUROSTOXX 50 & 2003-2020 \\
MW & WHEAT, MINN & 1990-2020 & XX & DOW JONES STOXX 50 & 2004-2020 \\
NR & NATURAL GAS & 1990-2020 & YM & DOW JONES, MINI (\$5.00) & 2004-2020 \\
SB & SUGAR \#11 & 1990-2020  & \multicolumn{2}{l}{\textbf{Fixed Income:}  }  \\
W\_ & WHEAT, CBOT & 1990-2018 & AP &  AUSTRALIAN PRICE INDEX & 2010-2020 \\
ZA & PALLADIUM, electronic &1990-2020  & DT &  EURO BOND (BUND) & 1991-2020\\
ZB & RBOB, electronic & 1990-2020 & FB &  T-NOTE, 5yr composite &1990-2020\\
ZC & CORN, electronic & 1990-2020 & GS &  GILT, LONG BOND & 1991-2020\\
ZF & FEEDER CATTLE, electronic & 1990-2020 & TU &  T-NOTES, 2yr composite & 1992-2020\\
ZG & GOLD, electronic & 1990-2020 & TY &  T-NOTE, 10yr composite & 1990-2020\\
ZI & SILVER, electronic &1990-2020 & UB &  EURO BOBL & 2000-2020\\
ZK & COPPER, electronic & 1990-2020 & US &  T-BONDS, composite & 1990-2020 \\
ZL & SOYBEAN OIL, electronic &1990-2020 & \multicolumn{2}{l}{\textbf{Currencies:}  }  \\
ZM & SOYBEAN MEAL, electronic &1990-2020 & AN & AUSTRALIAN \$\$, day session & 1990-2020 \\
ZN & NATURAL GAS, electronic & 1992-2020 & BN & BRITISH POUND, composite & 1990-2020\\
ZO & OATS, electronic &1990-2020  & CB & CANADIAN 10YR BOND & 1996-2020\\
ZP & PLATINUM, electronic & 1990-2020 & CN & CANADIAN \$\$, composite & 1990-2020\\
ZR & ROUGH RICE, electronic & 1990-2020 & DX & US DOLLAR INDEX & 1990-2020\\
ZS & SOYBEANS electronic & 1990-2020 & FN & EURO, composite &1990-2020\\
ZT & LIVE CATTLE, electronic & 1990-2020 & JN & JAPANESE YEN, composite & 1990-2020\\
ZU & CRUDE OIL, electronic & 1990-2020 & MP & MEXICAN PESO & 1997-2020\\
ZW & WHEAT electronic & 1990-2020 & NK & NIKKEI INDEX & 1992-2020\\
ZZ & LEAN HOGS, electronic & 1990-2020 & SN & SWISS FRANC, composite &1990-2020 \\

\bottomrule
\end{tabular}
\end{table*}

The backtest was primarily conducted with the Pinnacle Data Corp CLC Database. We have access to a robust dataset comprising 98 ratio-adjusted continuous futures contracts dating back to 1990. However, for the purpose of this study, we narrowed our focus to a subset of 64 assets, as listed in Table \ref{table:universe}. \par

\end{document}